\documentclass[reprint,superscriptaddress,amsmath,amssymb,aps,prc,twocolumn,floatfix]{revtex4-2}

\usepackage{graphicx}
\usepackage{hyperref}
\usepackage{xcolor}
\usepackage[english]{babel}
\usepackage{bm}
\usepackage[T1]{fontenc}
\usepackage{multirow}
\usepackage[utf8]{inputenc}

\usepackage{lineno}

\usepackage[output-decimal-marker={.},exponent-product=\cdot]{siunitx}

\begin{document}

\title{High-precision mass measurements of the ground and isomeric states in $^{124,125}$Ag}

\author{J. Ruotsalainen}
\email{jouni.k.a.ruotsalainen@jyu.fi}
\affiliation{University of Jyvaskyla, Department of Physics, Accelerator laboratory, P.O. Box 35(YFL) FI-40014 University of Jyvaskyla, Finland}
\author{D.~A.~Nesterenko}
\affiliation{University of Jyvaskyla, Department of Physics, Accelerator laboratory, P.O. Box 35(YFL) FI-40014 University of Jyvaskyla, Finland} 
\author{M. Stryjczyk}
\email{marek.m.stryjczyk@jyu.fi}
\affiliation{University of Jyvaskyla, Department of Physics, Accelerator laboratory, P.O. Box 35(YFL) FI-40014 University of Jyvaskyla, Finland} 
\author{A.~Kankainen}
\email{anu.kankainen@jyu.fi}
\affiliation{University of Jyvaskyla, Department of Physics, Accelerator laboratory, P.O. Box 35(YFL) FI-40014 University of Jyvaskyla, Finland}
\author{L.~Al~Ayoubi}
\affiliation{University of Jyvaskyla, Department of Physics, Accelerator laboratory, P.O. Box 35(YFL) FI-40014 University of Jyvaskyla, Finland}
\affiliation{Universit\'e Paris Saclay, CNRS/IN2P3, IJCLab, 91405 Orsay, France}
\author{O.~Beliuskina}
\affiliation{University of Jyvaskyla, Department of Physics, Accelerator laboratory, P.O. Box 35(YFL) FI-40014 University of Jyvaskyla, Finland}
\author{L.~Canete}
\altaffiliation[Present address: ]{Northeastern University London, Devon House, 58 St Katharine's Way, E1W 1LP, London, United Kingdom}
\affiliation{University of Jyvaskyla, Department of Physics, Accelerator laboratory, P.O. Box 35(YFL) FI-40014 University of Jyvaskyla, Finland}
\author{P.~Chauveau}
\affiliation{CSNSM, bat. 101, Domaine de l'Université de Paris Sud, 91400 Orsay, France}
\author{R.~P.~de Groote}
\altaffiliation[Present address: ]{KU Leuven, Instituut voor Kern- en Stralingsfysica, B-3001 Leuven, Belgium}
\affiliation{University of Jyvaskyla, Department of Physics, Accelerator laboratory, P.O. Box 35(YFL) FI-40014 University of Jyvaskyla, Finland}
\author{P.~Delahaye}
\affiliation{GANIL, CEA/DSM-CNRS/IN2P3, Boulevard Henri Becquerel, 14000 Caen, France}
\author{T.~Eronen}
\affiliation{University of Jyvaskyla, Department of Physics, Accelerator laboratory, P.O. Box 35(YFL) FI-40014 University of Jyvaskyla, Finland}
\author{M.~Flayol}
\affiliation{Universit\'e de Bordeaux, CNRS/IN2P3, LP2I Bordeaux, UMR 5797, F-33170 Gradignan, France}
\author{Z.~Ge}
\affiliation{GSI Helmholtzzentrum für Schwerionenforschung, 64291 Darmstadt, Germany}
\affiliation{University of Jyvaskyla, Department of Physics, Accelerator laboratory, P.O. Box 35(YFL) FI-40014 University of Jyvaskyla, Finland}
\author{S.~Geldhof}
\altaffiliation[Present address: ]{GANIL, CEA/DRF-CNRS/IN2P3, B.P. 55027, 14076 Caen, France}
\affiliation{University of Jyvaskyla, Department of Physics, Accelerator laboratory, P.O. Box 35(YFL) FI-40014 University of Jyvaskyla, Finland}
\author{W.~Gins}
\affiliation{University of Jyvaskyla, Department of Physics, Accelerator laboratory, P.O. Box 35(YFL) FI-40014 University of Jyvaskyla, Finland}
\author{M.~Hukkanen}
\affiliation{University of Jyvaskyla, Department of Physics, Accelerator laboratory, P.O. Box 35(YFL) FI-40014 University of Jyvaskyla, Finland}
\affiliation{Universit\'e de Bordeaux, CNRS/IN2P3, LP2I Bordeaux, UMR 5797, F-33170 Gradignan, France}
\author{A.~Jaries}
\affiliation{University of Jyvaskyla, Department of Physics, Accelerator laboratory, P.O. Box 35(YFL) FI-40014 University of Jyvaskyla, Finland}
\author{D.~Kahl}
\altaffiliation[Present address: ]{Facility for Rare Isotope Beams, Michigan State University, East Lansing, MI, USA}
\affiliation{Extreme Light Infrastructure – Nuclear Physics, Horia Hulubei National Institute for R\&D in Physics and Nuclear Engineering (IFIN-HH), 077125 Bucharest-M\u{a}gurele, Romania}
\author{D.~Kumar}
\affiliation{GSI Helmholtzzentrum für Schwerionenforschung, 64291 Darmstadt, Germany}
\author{I.~D.~Moore}
\affiliation{University of Jyvaskyla, Department of Physics, Accelerator laboratory, P.O. Box 35(YFL) FI-40014 University of Jyvaskyla, Finland}
\author{S.~Nikas}
\affiliation{University of Jyvaskyla, Department of Physics, Accelerator laboratory, P.O. Box 35(YFL) FI-40014 University of Jyvaskyla, Finland} 
\author{H.~Penttil\"a}
\affiliation{University of Jyvaskyla, Department of Physics, Accelerator laboratory, P.O. Box 35(YFL) FI-40014 University of Jyvaskyla, Finland}
\author{D.~Pitman-Weymouth}
\affiliation{Department of Physics and Astronomy, University of Manchester, Manchester M13 9PL, United Kingdom}
\author{A.~Raggio}
\affiliation{University of Jyvaskyla, Department of Physics, Accelerator laboratory, P.O. Box 35(YFL) FI-40014 University of Jyvaskyla, Finland}
\author{S.~Rinta-Antila}
\affiliation{University of Jyvaskyla, Department of Physics, Accelerator laboratory, P.O. Box 35(YFL) FI-40014 University of Jyvaskyla, Finland}
\author{A.~de Roubin}
\altaffiliation[Present address: ]{LPC Caen, Normandie Univ., 14000 Caen, France}
\affiliation{University of Jyvaskyla, Department of Physics, Accelerator laboratory, P.O. Box 35(YFL) FI-40014 University of Jyvaskyla, Finland}
\author{M.~Vilen}
\affiliation{University of Jyvaskyla, Department of Physics, Accelerator laboratory, P.O. Box 35(YFL) FI-40014 University of Jyvaskyla, Finland}
\author{V.~A.~Virtanen}
\affiliation{University of Jyvaskyla, Department of Physics, Accelerator laboratory, P.O. Box 35(YFL) FI-40014 University of Jyvaskyla, Finland}
\author{M.~Winter}
\affiliation{University of Jyvaskyla, Department of Physics, Accelerator laboratory, P.O. Box 35(YFL) FI-40014 University of Jyvaskyla, Finland}

\begin{abstract}
The masses of the ground and isomeric states in $^{124,125}$Ag have been measured using the phase-imaging ion-cyclotron-resonance technique at the JYFLTRAP double Penning trap mass spectrometer. The ground states of $^{124}$Ag and $^{125}$Ag were found to be 30(250)~keV and 250(430)~keV less bound but 36 and 110 times more precise than in the Atomic Mass Evaluation 2020, respectively. The excitation energy of $^{124}$Ag$^{m}$, ${E_x = 188.2(25)}$~keV, was determined for the first time. The new precise mass values have been utilised to study the evolution of nuclear structure via two-neutron separation energies. The impact on the astrophysical rapid neutron capture process has been investigated via neutron-capture reaction rate calculations. The precision measurements indicate a more linear trend in two-neutron separation energies and reduce the mass-related uncertainties for the neutron-capture rate of $^{124}$Ag$(n,\gamma)^{125}$Ag by a factor of around 100. The new mass values also improve the mass of $^{123}$Pd, previously measured using $^{124}$Ag as a reference. 
\end{abstract}

\maketitle

\section{Introduction}

The region of the doubly-magic $^{132}$Sn isotope is interesting from the point of view of nuclear structure and nuclear astrophysics studies. The rapid neutron capture process ($r$ process) \cite{Burbidge1957}, which is responsible for the production of about 50\% of the elemental abundances above iron \cite{Cowan2021}, leads to a strong production of isotopes around $A\sim130$, creating a so-called second $r$-process peak. Sensitivity studies have shown that the nuclear properties close to $^{132}$Sn have the highest impact on the $r$-process abundances for different astrophysical scenarios \cite{Mumpower2015,Mumpower2016}.

Several low-lying isomers in the region of $^{132}$Sn, especially for neutron numbers ${N<82}$, have been observed \cite{NUBASE20}. Their presence allows for studies of shell evolution \cite{Watanabe2024} or isomeric yield ratios \cite{Rakopoulos2019,Gao2023}. The properties of metastable states are also relevant for astrophysical calculations \cite{Misch2021,Misch2024}. However, the isomerism results in an additional difficulty in measuring nuclear properties such as masses, half-lives and beta-delayed neutron emission probabilities $P_{xn}$, crucial from the $r$-process point of view, as they are often challenging to identify and resolve from the ground state. Consequently, the extracted results can often be averages of the values of each long-lived state, mixed in an unknown proportion. This phenomenon has been observed in the extraction of, for example, the $P_{1n}$ values and half-lives of the states in $^{131}$In \cite{Phong2022,Dunlop2019,Benito2024} or masses of the states in the odd-N rhodium and ruthenium isotopes \cite{Hager2007,Hukkanen2023,Hukkanen2023a}.

One of the isotopic chains known to host several isomers is silver (${Z=47}$). While the masses of these isotopes were measured up to $^{126}$Ag \cite{Breitenfeldt2010,Knoebel2016}, the utilised techniques did not provide enough resolving power to separate ground states from the long-lived isomers. As a result, the isomer excitation energies of $^{120,124}$Ag are currently not known \cite{NUBASE20} while in the case of $^{122}$Ag one of the proposed isomers was determined to be non-existent \cite{Jaries2024a}. In addition, the uncertainty of the mass value of $^{125}$Ag was increased by the Atomic Mass Evaluation 2020 (AME20) evaluators by a factor 2.5 while the result for $^{126}$Ag was rejected as it was deemed unreliable \cite{Huang2021}. 

With the advent of the phase-imaging ion-cyclotron-resonance (PI-ICR) technique \cite{Eliseev2014,Nesterenko2018} implemented at Penning traps, it has become possible to resolve isomeric states separated as little as 10 keV \cite{Huang2021,Orford2018}. This method has been already successfully applied to other silver isotopes \cite{deGroote2024,Jaries2024a,Ge2024}. Extraction of pure ground-state masses allowed for comparison with different theoretical models which resulted in a better understanding of the underlying nuclear structure.

In this work, we report the first high-precision mass measurements of the long-living states in $^{124,125}$Ag using the PI-ICR technique at the JYFLTRAP double Penning trap \cite{Eronen2012}. We also report the first measurement of the $^{124}$Ag$^{m}$ excitation energy. The results are compared with six density-functional-theory-based theoretical models, commonly used in predictions of nuclear structure and astrophysical calculations. Additionally, the impact of the measurements on the mass of $^{123}$Pd and the astrophysical reaction rates are discussed.

\section{Experimental methods}

The experiment was performed at the Ion Guide Isotope Separator On-Line (IGISOL) facility in Jyv\"askyl\"a, Finland \cite{Moore2013,Penttila2020}. The radioactive ion beam was produced in proton-induced fission by impinging a 25-MeV proton beam into a 15~mg/cm$^2$-thick $^{nat}$U target. The fission products were stopped in a helium-filled gas cell operated at about 300~mbar. Then, the ions were guided out by the gas flow into a sextupole ion guide, accelerated by a 30 kV potential difference and purified with respect to their mass-over-charge ratio $m/q$ by a 55$^\circ$ dipole magnet. Subsequently, the beam was injected into a helium-filled radio-frequency quadrupole cooler-buncher \cite{Nieminen2001}. From there, the bunched beam was delivered to the JYFLTRAP double Penning trap mass spectrometer \cite{Eronen2012}. 

At JYFLTRAP, the ions were first cooled, purified and centered in the first (preparation) trap by using a mass-selective buffer-gas cooling technique \cite{Savard1991}. The  singly-charged ions of interest were sent to the second (measurement) trap and back to the first trap for an additional cooling stage before their cyclotron frequency $\nu_c = qB/(2 \pi m)$ in the magnetic field $B$ was determined using the PI-ICR technique \cite{Eliseev2013,Eliseev2014,Nesterenko2018,Nesterenko2021}. 

In the PI-ICR method, the phases of the radial eigenmotions after an accumulation time $t_{acc}$ are determined by projecting the trapped ions onto a position-sensitive microchannel-plate detector (2D MCP). Using the polar angles of the magnetron ($\alpha_-$) and reduced cyclotron ($\alpha_+$) phase spots with respect to the trap center on the 2D MCP, the cyclotron frequency is obtained as:
\begin{equation}
    \nu_c=\frac{\alpha_c+2\pi n_c}{2\pi t_\text{acc}} \mathrm{,}
\end{equation}
where $n_c$ denotes the number of revolutions the ion would perform in magnetic field during $t_{acc}$, while $\alpha_c = \alpha_+ - \alpha_-$ is the difference between polar angles of the phase spots. 

\begin{figure}[htb]
    \centering
    \includegraphics[width=\columnwidth]{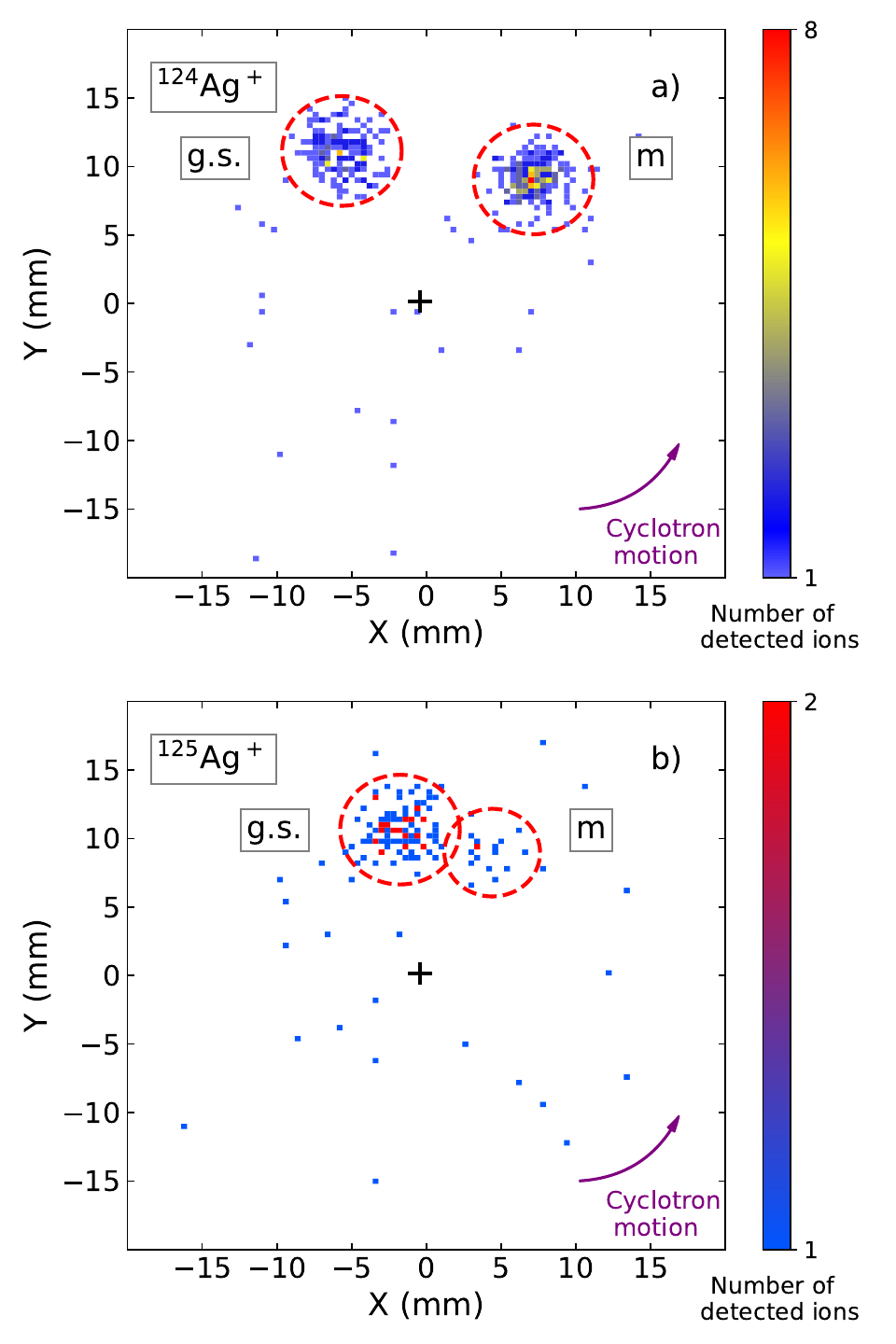}
    \caption{\label{fig:piicr} Projections of the ions' cyclotron motions onto the 2D MCP detector obtained with the PI-ICR technique for a) $^{124}$Ag$^+$ ions with the accumulation time ${t_{\rm acc} = 130}$~ms and b) $^{125}$Ag$^+$ ions with the accumulation time of ${t_{\rm acc} = 116}$~ms. The position of the center spots are indicated with the $+$ symbol.}
\end{figure}

Typical PI-ICR images obtained in this work for $^{124}$Ag and $^{125}$Ag are shown on Fig.~\ref{fig:piicr}. A phase accumulation time of ${t_{\rm acc}=130}$~ms was applied for $^{124}$Ag and ${t_{\rm acc}=116}$~ms for $^{125}$Ag. With these accumulation times the ground and isomeric states could be resolved from each other (see Fig.~\ref{fig:piicr}).

For $^{124}$Ag, an ion entering the cooler-buncher at the start of the measurement cycle reached the 2D-MCP after 482 ms. However, ion bunches were extracted from the measurement trap every 292 ms, as the cooler-buncher and the preparation and measurement traps could be operated simultaneously. Similarly, a $^{125}$Ag ion entering the cooler at the start of the cycle took 334 ms to reach the detector, but the measurement cycle length was 226 ms.

The magnetic field strength $B$ was determined using $^{133}$Cs$^+$ ions ($\Delta_{\rm lit.} = -88070.943(8)$ keV \cite{AME20}) produced in an offline surface ion source \cite{Vilen2020}. The atomic mass $M$ is determined from the cyclotron frequency ratio $r=\nu_{c,{\rm ref}}/\nu_{c}$ between the reference ion and the ion of interest:
\begin{equation}\label{eq:mass}
M = (M_{\rm ref} - m_{e}) r + m_{e}\mathrm{,}
\end{equation}
where $M_{\rm ref}$ and $m_{e}$ are the atomic mass of the reference ion and the electron mass, respectively. The $^{124}$Ag isomer excitation energy $E_x$ was extracted using the mass of $^{124}$Ag$^{gs}$ as a reference:
\begin{equation}
E_x = (r-1)[M_{\rm ref} - m_e]c^2 \mathrm{,}
\end{equation} 
with $c$ being the speed of light in vacuum. 

To account for the temporal fluctuations of the magnetic field, which results in a standard uncertainty of $\delta B/(B\delta t) = 2.01\times10^{-12}$ min$^{-1}$ \cite{Nesterenko2021}, the measurements of the ion of interest and the reference ion were alternated. For the $^{124}$Ag measurement, the ion of interest was collected for about seven minutes, followed by about two minutes of the reference ion. In case of the $^{124}$Ag$^{m}$ excitation energy both species were alternated every two minutes. For both states in $^{125}$Ag, the silver ions were measured for about 25 minutes, followed by three minutes of $^{133}$Cs. The electron binding energies in these cases are negligible and were therefore not taken into account in the calculations. A mass dependent systematic uncertainty of $2.3\times10^{-10}/u\times\Delta m$, where $\Delta m$ is the mass difference of the reference and the ion of interest, and residual uncertainty of $5.3\times 10^{-9}$, as reported in Ref.~\cite{Nesterenko2021}, were included to the cyclotron frequency ratios $r$ in the measurements using $^{133}$Cs as a reference. In the case of $^{124}$Ag, the count-rate class analysis \cite{Roux2013} with up to three ions per bunch was also performed to take into account ion-ion interactions in the measurement trap. 

\section{Results}

\begin{table*}
\centering
\begin{ruledtabular}
\caption{\label{tab:results}The mass measurement results of the nuclides studied in this work together with their spin-parities $J^{\pi}$ and half-lives $T_{1/2}$ from literature \cite{Batchelder2014,Batchelder2021}. Columns Ref. and $r=\nu_{c,{\rm ref}}/\nu_{c}$ show the reference ions and the measured cyclotron frequency ratios, respectively. Corresponding mass-excess values $\Delta$ and isomer excitation energies $E_x$ are tabulated and compared to the literature values $\Delta_{\rm lit.}$ and $E_{x,{\rm lit.}}$ from Refs. \cite{AME20,NUBASE20}. The differences between this work and the literature $\mathrm{Diff.} = \Delta - \Delta_{\rm lit.}$ are also added. \# indicates an extrapolated value.}
\begin{tabular}{llllllllll}
Nuclide & $J^{\pi}$ & $T_{1/2}$ (ms) & Ref. & $r=\nu_{c,{\rm ref}}/\nu_{c}$ & $\Delta$ (keV) & $\Delta_{\rm lit.}$ (keV)  & $E_x$ (keV) & $E_{x,{\rm lit.}}$ (keV) & Diff. (keV)\\\hline
$^{124}$Ag          & $(2^-)$   & 191(28) & $^{133}$Cs   & \num{0.932 459 075(56)}  & \num{-66199.6(69)} 	& \num{-66230(250)}    &              &  & 30(250)\\
$^{124}$Ag$^{m}$    & $(8^-)$   & 144(20) & $^{124}$Ag   & \num{1.000 001 631(22)}  & \num{-66011.4(73)} 	& \num{-66180(260)}\# & 188.2(25)    & 50(50)\# & 169(260)\#\\
$^{125}$Ag          & $(9/2^+)$   & 176(3) & $^{133}$Cs  & \num{0.939 998 835(33)}  & \num{-64270.4(40)}    & \num{-64520(430)}     &              &  & 250(430)\\
$^{125}$Ag$^{m}$    & $(1/2^-)$   & 159(21) & $^{133}$Cs & \num{0.939 999 716(64)}	& \num{-64161.4(79)}	& \num{-64420(430)} & 109.0(89)  & 97.1(5) & 259(430)\\
\end{tabular}
\end{ruledtabular}
\end{table*}

The results of this work are summarized in Table~\ref{tab:results} and compared to the literature values given in the Atomic Mass Evaluation 2020 (AME20) \cite{AME20} for the ground states and NUBASE20 for the isomers \cite{NUBASE20}. All mass-excess values $\Delta=(M-Au)c^2$, where $A$ is the atomic mass number and $u$ is the atomic mass unit, were determined with a precision better than 8 keV.

The measured mass-excess value of $^{124}$Ag, ${\Delta = -66199.6(69)}$~keV, agrees with the ISOLTRAP result (${\Delta_{\rm lit.} = -66200(250)}$~keV \cite{Breitenfeldt2010}), however, it is 36 times more precise. It should be noted that the AME20 value (${\Delta_{\rm lit.} = -66230(250)}$~keV \cite{AME20}) is shifted by 30~keV with respect to the ISOLTRAP value to correct for the presence of the isomeric state which was estimated at $50(50)\#$~keV \cite{Huang2021}. The isomer excitation energy, $E_x = 188.2(25)$~keV, was measured for the first time and it is 138(50)\#~keV larger than the NUBASE20 extrapolation \cite{NUBASE20}.

It is not well established in literature \cite{NUBASE20,ensdf}, which of the two long-lived states of $^{124}$Ag is the ground state. To identify the states, the measured isomeric yield ratio was studied. The ratio of the number of measured ions of the isomer to the ground state is about $1.5:1$. If the ground state is assumed to be the low-spin longer-lived state and the isomer high-spin and shorter lived, this ratio would be about $2.5:1$ at the moment of production, while for the reversed order, the ratio would become about $1:1$. The facts that fission favors the production of high-spin states \cite{Rakopoulos2019,Gao2023,Nesterenko2023} and that in less exotic odd-odd Ag isotopes isomers are the high-spin states \cite{NUBASE20,Jaries2024a} suggests the order proposed in the literature \cite{NUBASE20} is correct.  

The measured mass of $^{125}$Ag, ${\Delta = -64270.4(40)}$~keV, agrees with the AME20 value of ${\Delta_{\rm lit.} = -64520(430)}$~keV \cite{AME20}, however, it is 110 times more precise. It should be noted that the AME20 value originates from an Experimental Storage Ring (ESR) measurement at GSI \cite{Knoebel2016} and the uncertainties were inflated by the evaluators by a factor 2.5, from 173 to 430~keV \cite{Huang2021}. In this work, the low-lying isomer was resolved from the ground state for the first time, see Fig.~\ref{fig:piicr}b. Its excitation energy, $E_x = 109.0(89)$~keV, is 11.9(89)~keV (1.3$\sigma$) larger than the precise value known from the $\beta$-decay study \cite{Chen2019}. The observed ground-state-to-isomer ratio of the number of ions at the 2D MCP after the phase accumulation time (about $19:1$) supports the literature assignment of the ground state as the high-spin state \cite{NUBASE20}.

\section{Discussion}

Differences in the mass trends are associated with changes of the ground-state nuclear structure, such as shell closures or an onset of deformation \cite{Lunney2003,Eronen2016,Garrett2022}. The effect of the mass values reported in this work on these mass trends in the neutron-rich silver isotopes was studied by analyzing the two-neutron separation energies $S_{2n}$ defined as:
\begin{equation}
S_{2n}(Z,N) = \Delta(Z,N-2) - \Delta(Z,N) +2\Delta_n \mathrm{,}
\end{equation}
where $\Delta(Z,N)$ is the mass excess of a nucleus with given proton ($Z$) and neutron ($N$) numbers and $\Delta_n$ is the mass excess of a free neutron. In addition, the two-neutron shell-gap energies $\delta_{2n}$:
\begin{equation}
\delta_{2n}(Z,N) = S_{2n}(Z,N) - S_{2n}(Z,N+2) \mathrm{,}
\end{equation}
were also determined. 

\begin{figure}
    \centering
    \includegraphics[width=\columnwidth]{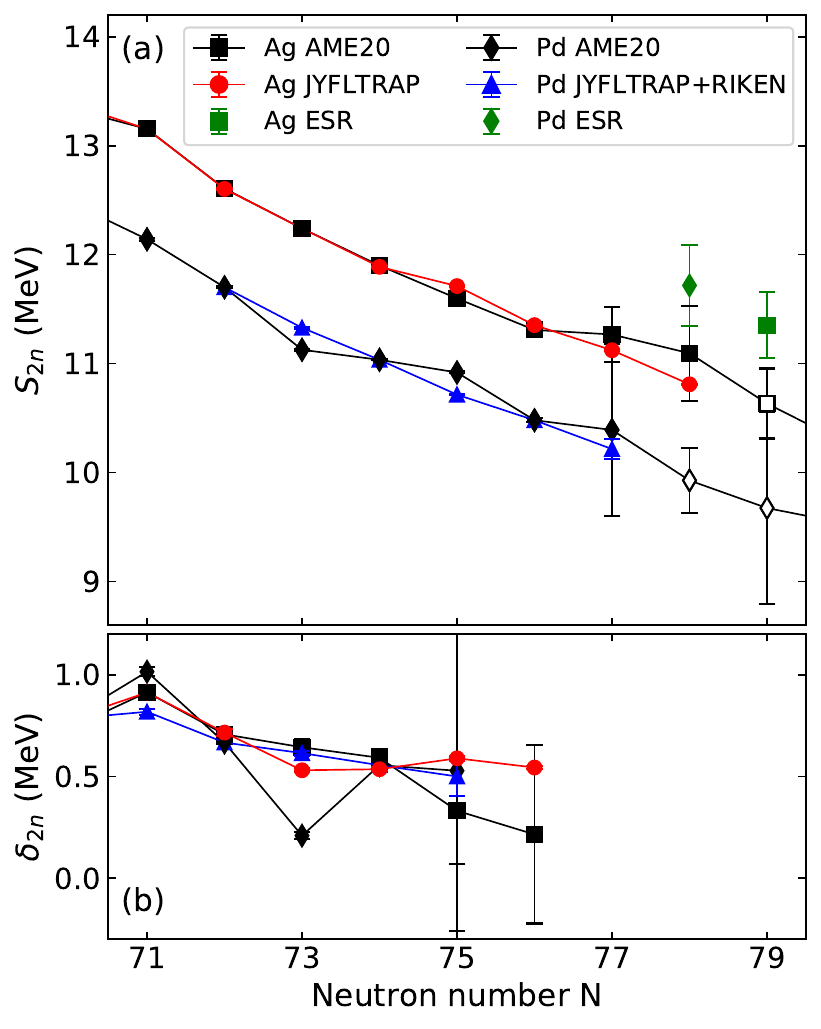}
    \caption{\label{fig:AgPdmasstrends}Two-neutron separation energies $S_{2n}$ (panel a) and two-neutron shell-gap energies $\delta_{2n}$ (panel b) for the neutron-rich Ag and Pd isotopes. The black squares and diamonds are calculated using the AME20 values \cite{AME20}, the red circles are taken from this work for $^{124,125}$Ag, from Ref. \cite{deGroote2024} for $^{117,119,121,123}$Ag and from Ref. \cite{Jaries2024a} for $^{122}$Ag, while the blue triangles are obtained using the redetermined mass of $^{123}$Pd \cite{Li2022} based on the mass of $^{124}$Ag from this work and the mass of $^{119}$Pd from Ref. \cite{Jaries2024a}. The green square and diamond indicates the $S_{2n}$ values calculated using the ESR mass values of $^{126}$Ag and $^{124}$Pd, respectively, reported in Ref. \cite{Knoebel2016} and rejected by the AME20 evaluators \cite{Huang2021}. Empty squares indicate the values based on the AME20 extrapolations \cite{AME20}.}
\end{figure}

Figure~\ref{fig:AgPdmasstrends} shows the two-neutron separation energies and shell-gap energies for Pd ($Z$=46) and Ag ($Z$=47) isotopic chains based on this work and recent JYFLTRAP publications \cite{deGroote2024,Jaries2024a}. For comparison, AME20 values are also plotted.  

While the refined $S_{2n}$ values from this work agree with the AME20 evaluation \cite{AME20} within one standard deviation, the new precise mass values verify the linearity of the $S_{2n}$ values in the region and reduce the probability for deviations, see Fig.~\ref{fig:AgPdmasstrends}a. This is more visible in the $\delta_{2n}$ values, which remain almost constant at around 530-590~keV for $73 \leq N \leq 76$ (see Fig.~\ref{fig:AgPdmasstrends}b). We note that our mass values shift the $S_{2n}$ trend away from the point based on the mass of $^{126}$Ag reported in Ref.~\cite{Knoebel2016}, which provides an additional support for the AME20 evaluators decision to exclude them.

\begin{figure}
    \centering
    \includegraphics[width=\columnwidth]{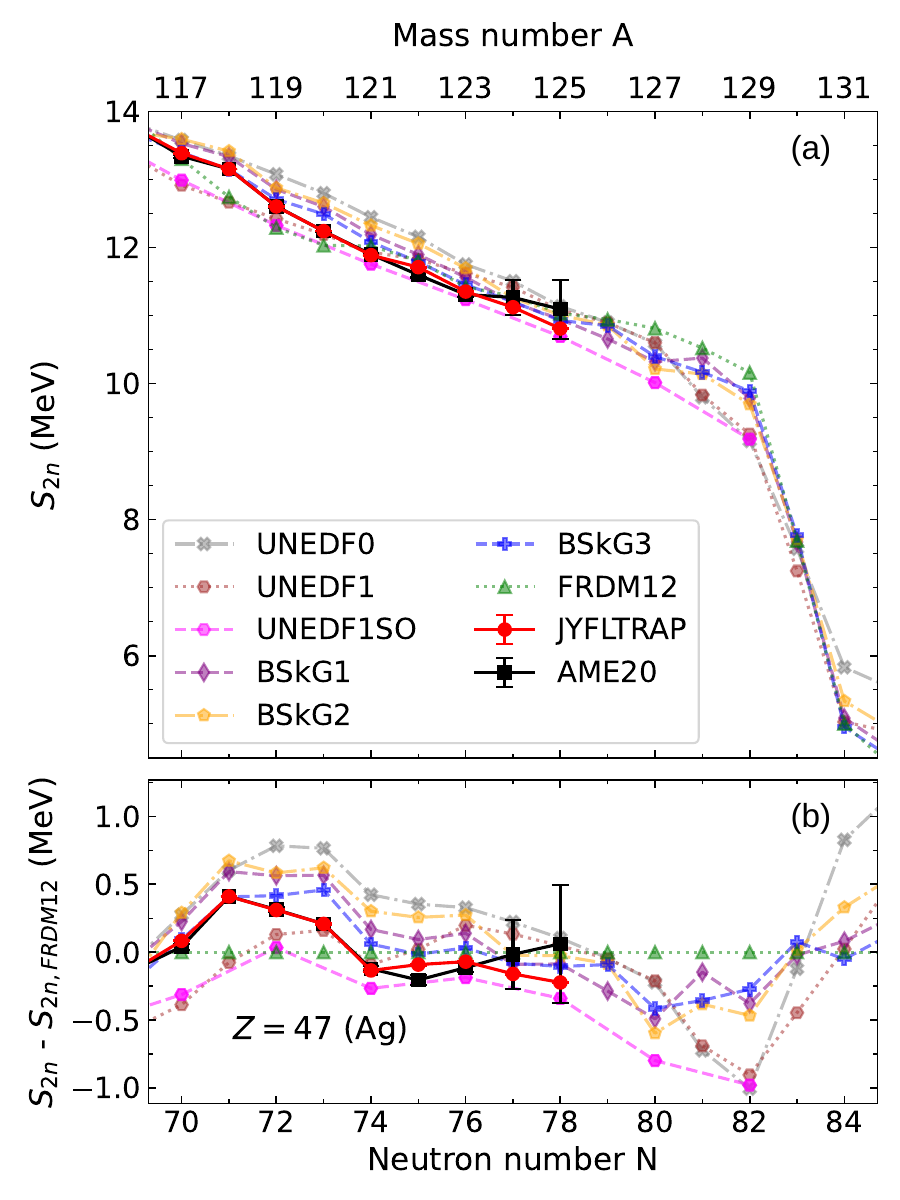}
    \caption{\label{fig:S2ntheory}Comparison of the experimental results for the Ag (${Z=47}$) isotopes to mass models UNEDF0 \cite{Kortelainen2010}, UNEDF1 \cite{Kortelainen2012}, UNEDF1$_{\text{SO}}$ \cite{Shi2014}, BSkG1 \cite{Scamps2021}, BSkG2 \cite{Ryssens2022}, BSkG3 \cite{Grams2023}, and FRDM12 \cite{Moeller2016}. (a) Two-neutron separation energies $S_{2n}$ and (b) differences between two-neutron separation energies from this work and other JYFLTRAP measurements \cite{deGroote2024,Jaries2024a}, AME20, and the theoretical models from the values given by the model FRDM12 ($S_{2n,FRDM12}$).}
\end{figure}

The experimental $S_{2n}$ values were compared to six mass models commonly used for interpreting nuclear structure and for astrophysical applications, see e.g. Refs.~\cite{Martin2016,Mumpower2016,Scamps2021}. Five of them are Density Functional Theory based: UNEDF0 \cite{Kortelainen2010}, UNEDF1 \cite{Kortelainen2012}, BSkG1 \cite{Scamps2021}, BSkG2 \cite{Ryssens2022} and BSkG3 \cite{Grams2023} while the last one is the FRDM12 model which is based on the combination of a finite-range liquid-drop model with a folded-Yukawa single-particle potential \cite{Moeller2016}. The results of this comparison are presented in Fig.~\ref{fig:S2ntheory}. To make the differences between the experimental values and the models more visible, we have plotted them as compared to the FRDM12 in Fig.~\ref{fig:S2ntheory}.b.

While all the models from the BSkG* family overestimate the $S_{2n}$ values for $N\geq73$ by about 200-300 keV, they very well reproduce the slope of the curve. Similarly, the UNEDF0 model overestimates the $S_{2n}$ values but the slope follows the experimental trend rather well. The slope predicted by the FRDM12 and UNEDF1 models in the $N\leq80$ region is less steep than predicted by the other models. All the used models work relatively well within the experimentally known region but start to deviate when approaching the shell closure at $N=82$, with deviations up to 1 MeV observed at $N=82$. This highlights the need for new experimental values of exotic silver isotopes.

We note that the UNEDF1 model failed to reproduce the excitation energies of the low-lying isomers in the even-$N$ $^{103-123}$Ag isotopes \cite{deGroote2024}, but the discrepancy was significantly reduced by using a modified UNEDF1$_{\text{SO}}$ functional with an increased spin-orbit strength \cite{Shi2014}. Also for the $S_{2n}$ values, the UNEDF1$_{\text{SO}}$ model follows the experimental trend better than UNEDF1 (see Fig.~\ref{fig:S2ntheory}), however, the discrepancy to the experimental mass-excess values increases with UNEDF1$_{\text{SO}}$. This further supports the need for improvements in the models.

With the new precise mass value of $^{124}$Ag, the mass of $^{123}$Pd reported in Ref.~\cite{Li2022} can be re-evaluated. Both $^{124}$Ag and $^{123}$Pd are known to have two long-lived states \cite{NUBASE20,Chen2019} and it is unclear which ones were measured in Ref.~\cite{Li2022}, see Tab.~\ref{tab:results_AME20_improved}. Since the production method of $^{124}$Ag from Ref.~\cite{Hall2021} is identical to Ref.~\cite{Li2022} and the reported half-life of $^{124}$Ag, ${T_{1/2} = 205(17)}$~ms \cite{Hall2021}, is consistent with the ground state (${T_{1/2} = 191(28)}$~ms \cite{Batchelder2021}) and not with the isomer (${T_{1/2} = 144(20)}$~ms \cite{Batchelder2021}), the $^{123}$Pd mass was redetermined using the mass-excess value of the $^{124}$Ag ground state. 

\begin{table}
\centering
\caption{\label{tab:results_AME20_improved}Mass excesses of the $N=77$ isotones from Ref. \cite{Li2022} compared to the values from AME20/NUBASE20 \cite{AME20,NUBASE20} and the redetermined values based on the new mass of $^{124}$Ag$^{gs}$ from this work. \#~indicates an extrapolated value.}
\begin{ruledtabular}
\begin{tabular}{lllll}
Isotope & $J^\pi$ & AME20 \cite{AME20} & Ref. \cite{Li2022} &  This work \\
& & (keV) & (keV) & (keV)\\\hline
$^{126}$In          & $3^+$ & $-77809(4)$ & \multirow{2}{*}{$-77707(266)$} &  \multirow{2}{*}{$-77677(90)$} \\
$^{126}$In$^m$      & $8^-$ & $-77719(5)$ & & \\\noalign{\vskip 1.3mm}
$^{125}$Cd          & $3/2^+$ & $-73348.1(29)$ & \multirow{2}{*}{$-73237(318)$} & \multirow{2}{*}{$-73207(196)$}  \\
$^{125}$Cd$^m$      & $11/2^-$ & $-73162(3)$ & & \\\noalign{\vskip 1.3mm}
$^{123}$Pd          & $3/2^+$\# & $-60430(790)$ & \multirow{2}{*}{$-60282(267)$} & \multirow{2}{*}{$-60253(95)$} \\
$^{123}$Pd$^m$      & $11/2^-$\# & $-60330(790)\#$ & & \\
\end{tabular}
\end{ruledtabular}
\end{table}

In the case of $^{126}$In and $^{125}$Cd, the updated mass values from Ref.~\cite{Li2022} agree with AME20. However, the large uncertainties do not allow for unambiguous determination which long-lived state was measured. For $^{126}$In, the redetermined value agrees with the isomeric-state mass and shows a $1.5\sigma$ deviation to the ground-state value, suggesting that the isomeric state has been more produced in Ref.~\cite{Li2022}. The refined mass value of $^{123}$Pd together with the new mass of $^{119}$Pd reported in Ref.~\cite{Jaries2024a} lead to a more linear trend in the $S_{2n}$ values for the Pd isotopes, see Fig.~\ref{fig:AgPdmasstrends}. We note that also the $\delta_{2n}$ curves for both isotopic chains, silver and palladium, are now following a similar trend. We also note that the updated mass value of $^{123}$Pd shifts the $S_{2n}$ trend even further away from the point based on the $^{124}$Pd mass reported in Ref.~\cite{Knoebel2016} which supports its rejection from AME20 \cite{Huang2021}.

\begin{figure}
    \centering
    \includegraphics[width=\columnwidth]{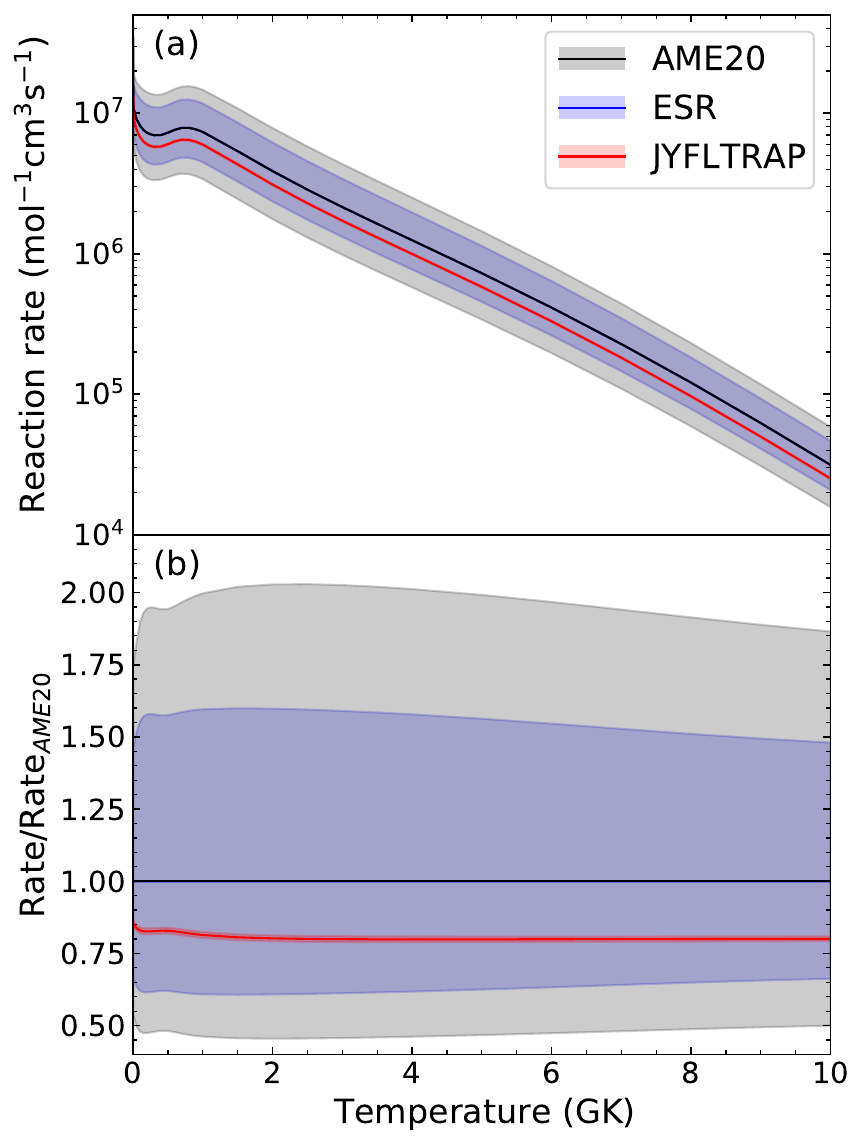}
    \caption{\label{fig:124125Ag_rr}(a) Astrophysical reaction rates calculated with the {\sc talys} 1.96 code \cite{Koning2023} for the reaction $^{124}$Ag$(n,\gamma)^{125}$Ag using the mass values from this work (JYFLTRAP, red), AME20 (black) \cite{AME20} and the $^{125}$Ag mass value from ESR (blue) \cite{Knoebel2016}. (b) Reaction rate ratios with respect to the AME20 rate.}
\end{figure}

The new mass values reported in this work result in a two orders of magnitude more precise neutron-capture $Q$-value for $^{124}$Ag, $Q_{n,\gamma}=6142(8)$~keV, with respect to the AME20 value of $Q_{n,\gamma}=6360(500)$~keV \cite{AME20}. Its impact on the astrophysical neutron-capture reaction rate of $^{124}$Ag$(n,\gamma)^{125}$Ag was studied using the {\sc talys} 1.96 code \cite{Koning2023}. A comparison between the results obtained with both mass values from either this work (JYFLTRAP) or AME20 \cite{AME20}, as well as the $^{124}$Ag value from AME20 combined with the mass of $^{125}$Ag from the ESR measurement with the original 173~keV uncertainty \cite{Knoebel2016}, is presented in Fig.~\ref{fig:124125Ag_rr}. The results are presented for $T\leq10$~GK which is relevant for neutron captures in the $r$ process, see e.g. Refs.~\cite{Cowan2021,Arnould2020}. While the reaction rate calculated with the masses from this work agree with both AME20 and ESR values, it is about 20\% lower and 100 times more precise. We have almost entirely removed the mass-related uncertainties in this reaction rate. 

The ${A\approx120}$ region shows very large variations in the calculated $r$-process abundances depending on the selected astrophysical conditions \cite{Kajino2017,MendozaTemis2015}. Moreover, the weak $r$~process in neutrino-driven winds of core-collapse supernovae can produce lighter $r$-process elements (${A\leq130}$) \cite{Arcones2014}, whereas the main $r$~process in dynamic ejecta of neutron-star mergers create dominantly heavier elements (${A\geq120}$) \cite{Martin2015}. Thus, both weak and main $r$ processes contribute to the abundances in the ${A\approx120}$ region and, therefore, the astrophysical conditions producing these nuclei vary significantly \cite{Martin2015}. We note that the solar-system $r$-process abundance uncertainties are also not negligible in this mass region \cite{Goriely1999,Cowan2021}. 

Since the astrophysical conditions and their individual contributions to the observed solar-system $r$-process abundances in this mass region are not well defined, comparisons of the solar abundances to the abundances or abundance ratios $Y_A(125)/Y_A(124)$ calculated with the newly determined reaction rates are not meaningful. Nevertheless, the new masses will help to pin down the abundances for cases where the astrophysical conditions are well constrained, or if the abundances or abundance ratios can be measured from an astrophysical site.

\section{Summary and outlook}

The ground-state masses of neutron-rich silver isotopes $^{124}$Ag and $^{125}$Ag were measured with high precision, reducing their mass uncertainties by a factor of 36 and 110, respectively, as compared to AME20. The isomeric states were resolved and their excitation energies determined via mass measurements. The excitation energy of the $^{124}$Ag isomer, ${E_x = 188.2(25)}$~keV, was reported for the first time in this work. The observed production ratios of long-lived states in both isotopes support the state ordering proposed in the literature. 

The updated masses, while in agreement with the previous AME20 values \cite{AME20}, resulted in more linear trends in two-neutron separation energies. The $^{124}\text{Ag}(n,\gamma)^{125}\text{Ag}$ reaction-rate calculations performed using the {\sc talys} 1.96 code \cite{Koning2023}, while within $1\sigma$ of the results with pre-existing data, indicated about 20\% decrease of reaction rate with two orders of magnitude reduction in mass-related uncertainties.

With the precise mass value of $^{124}$Ag from this work, the mass of $^{123}$Pd obtained using $^{124}$Ag as a reference in Ref.~\cite{Li2022}, was redetermined. The new, more precise value for $^{123}$Pd, together with the recent mass measurements of less-exotic palladium isotopes \cite{Jaries2024a}, resulted in a more linear $S_{2n}$ trend for the these isotopes, similar to the one observed in silver isotopes. 

A comparison of two-neutron separation energies with six commonly used mass models showed an overall good agreement between the experiment and theory. Although the $S_{2n}$ values predicted by the mass models agree rather well within the known region, they show a larger scatter toward the shell closure at $N=82$. To better understand the evolution of nuclear structure and the strength of the shell closure, studies of more exotic silver isotopes up to and crossing the ${N=82}$ magic number are needed. This is an experimental challenge as the half-lives approach the limits of Penning-trap measurements. While multi-reflection time-of-flight mass spectrometers have demonstrated their utility for mass measurements of exotic nuclei and their isomers (see e.g. Refs.~\cite{Hornung2020,Izzo2021,Nies2023}), the mass resolving power may not be sufficient to separate long-lived isomeric states in the more neutron-rich silver isotopes. For example, the first isomeric state both in $^{127}$Ag and $^{129}$Ag is predicted to have an excitation energy of around 20(20)~keV and a half-life on the order of 10-20~ms, and in more exotic silver isotopes the existence of isomeric states is unknown \cite{NUBASE20}. While the evolution of the $N=82$ shell gap can be studied even with a rather modest precision of around hundred keV, detailed understanding of nuclear structure and isomeric states around $N=82$ and beyond will require developments in experimental techniques. 

\begin{acknowledgments}

This project has received funding from the European Union’s Horizon 2020 research and innovation programme under grant agreements No. 771036 (ERC CoG MAIDEN) and No. 861198–LISA–H2020-MSCA-ITN-2019, and from the Research Council of Finland projects No. 295207, 306980, 327629, 354589 and 354968. J.R. acknowledges financial support from the Vilho, Yrj\"o and Kalle V\"ais\"al\"a Foundation. D.Ku. acknowledges the support from DAAD grant number 57610603.

\end{acknowledgments}

\bibliography{biblio}

\end{document}